\documentclass[12pt,a4paper]{article}
\usepackage{fullpage}
\usepackage{epsfig}
\usepackage{amssymb}
\usepackage{amstext}
\usepackage{amsmath}
\usepackage{cite}
\usepackage[english]{babel}
\usepackage{indentfirst}

\def\slantfrac#1#2{\hbox{$\,^{#1}\!/_{#2}$}}
\def\dfrac#1#2{{\displaystyle#1\over\displaystyle#2}}

\def\apgt{\ {\raise-.5ex\hbox{$\buildrel>\over{\scriptstyle\sim}$}}\ }
\def\aplt{\ {\raise-.5ex\hbox{$\buildrel<\over{\scriptstyle\sim}$}}\ }

\newcommand{\grad}{\mathop{\rm grad}\nolimits}

\renewcommand{\vec}[1]{{\bf #1}}
\renewcommand\frac\dfrac
\newcommand{\ddiv}{\mathop{\rm div}\nolimits}

\begin{document}

\title{Superhumps in Binary Systems and Their Connection to Precessional
Spiral Density Waves}

\author{P.V. Kaygorodov$^1$ , D.V. Bisikalo$^1$, O.A. Kuznetsov$^{1,2}$
   and A. A. Boyarchuk$^1$}
\date{}
\maketitle
\centerline{\small $^1$ Institute of Astronomy, Moscow, Russia}
\centerline{\small $^2$ Keldysh Institute of Applied Mathematics, Moscow, 
            Russia}

\begin{abstract}

We consider a mechanism for the formation of superhumps in the TV~Col system,
based on the possible existence of a precessional spiral wave in the accretion
disk of the system. This mechanism can act in binaries with arbitrary 
component-mass ratios, and our precessional spiral wave model can be applied
to explain observed superhumps of all types.

\end{abstract}

\section{Introduction}

Superhumps are modulations of the light curves of binary systems with periods
that differ from the orbital periods by several percent, and are observed
mainly during superoutbursts in SU~UMa systems. The main observational
features of superhumps are described in~\cite{Warner:95}.
Various authors have put forward models to explain the superhump phenomenon
(a brief description of various models and their problems are given,
for instance, in~\cite{Warner:95}). Currently, the most popular model explains
these light variations as an effect of precession of the outer regions of the
accretion disk. The presence of the Lindblad 3:1 resonance in the disk
results in an instability that leads to precession of the outer part of the
disk, with a period that is appreciably longer than the orbital period. The
beating of the orbital and precessional periods gives rise to the periodic
variations that are identified with superhumps. This model has several
shortcomings, the most important being that it implies a limit on the maximum
component mass ratio. In order for the Lindblad 3:1 resonance to be located
inside the accretion disk, the ratio of the donor and accretor masses $q$ must
be lower than $\sim 0.33$~\cite{Paczynski:77}, while there are several observed
systems with superhumps that have appreciably higher $q$ values, such as
TV~Col.

TV~Col is a ``permanent superhumper'', or one of a class of systems in which
superhumps are always observed. It was suggested in~\cite{Osaki:96} that
cataclysmic variables with high mass-transfer rates (TV~Col is such a system)
may persist in a superhump regime that is not interrupted by quiescent periods.
Among stars with superhumps, TV~Col has an unusually long orbital period,
$P_{orb}\sim 5.5$~h~\cite{Retter-et-al:2003}. Since there is a direct
correlation between the orbital period of a system and the donor mass 
(see, for instance,~\cite{Osaki:96}), this may be as high as 
$\sim 0.56 M_{\odot}$, while the component-mass ratio is probably in the range
$q\sim0.6\div 0.9$~\cite{Hellier:93}.It is obvious that the Lindblad 3:1
resonance cannot be located inside the accretion disk with such a $q$ value.
This casts doubt on the popular superhump model indicated above.

In 2004, a
new mechanism for the formation of superhumps in SU~UMa systems was suggested
in~\cite{AR:2004:PrecWave,AR:2004:SUUMa}. The basis of this mechanism is the
idea that a precessional-type density wave can form in the accretion disk.

The precessional wave in the accretion disks forms as a result of interactions
between elliptical streamlines. The asphericity of the gravitational field of
the binary system forces the semimajor axes of the streamlines to precess
opposite to the direction in which the matter flows. The rate of this
precession is proportional to the semimajor axis of the corresponding
streamline. Since the streamlines in the disk cannot intersect, their
interaction results in the establishment of a certain equilibrium precession
rate of all the streamlines, with their apastrons lining up so that they form
a spiral pattern. Because the matter velocity is minimum at the apastrons,
the density of the matter at these points grows, and a precessional spiral
density wave is formed in the disk.

After the formation of the precessional
density wave in the disk, the rate of accretion grows sharply (by up to an
order of magnitude). Matter approaches the surface of the accretor along the
precessional wave, and the region of accretion is localized in azimuth, and,
hence, forms a radiating spot at the surface of accretor. The increase in the
accretion rate due to the density wave explains both the development of a
superoutburst and the amplitude of the superhump. The wave slowly precesses
in the stationary (observer's) coordinate frame, leading to a shift of the
region of enhanced accretion (i.e., of the superhump) with each rotation of
the system. The beating of the orbital period and the precessional period of
the wave result in the superhump period, which is slightly larger than the
orbital period.

This superoutburst model based on the presence of a precessional spiral wave
in the accretion disk~\cite{AR:2004:SUUMa} made it possible for the first
time to explain all important observational manifestations of superoutbursts
and superhumps in SU~UMa systems, including their period, duration, and
energy, the anticorrelation of the brightness and color temperature at the
maximum of an ordinary superhump, the late-superhump phenomenon, etc.
Moreover, this model does not place strict constraints on the component mass
ratio, and so can be applied to the superhumps in the TV~Col system. 

The main aim of the present paper is to investigate the possible formation of
precessional spiral waves in accretion disks in systems with $q > 0.33$. We
will also consider the possibility of explaining all types of superhumps with
this model. 

\section{The Model}

The model used for our numerical simulations of mass flows in binary systems
is described in~\cite{AR:2003:CoolDiscsEng}. The flow is described by a
three-dimensional system of gravitational gas-dynamics equations, including
nonadiabatic radiative heating and cooling of the gas:

\begin{equation}
\left\{
\begin{array}{l}
\dfrac{\partial\rho}{\partial t}+\ddiv\rho\vec{v}=0\,,\\[3mm]
\dfrac{\partial\rho\vec{v}}{\partial t}
+\ddiv(\rho\vec{v}\otimes\vec{v})+\grad P=-\rho\grad\Phi
 - 2 [\vec{\Omega}_{bin}\times\vec{v}]\rho
\,,\\[5mm]
\dfrac{\partial\rho(\varepsilon+|\vec{v}|^2/2)}{\partial t}
+\ddiv\rho\vec{v}(\varepsilon+P/\rho+|\vec{v}|^2/2)=\\
\qquad\qquad-\rho\vec{v}\grad\Phi+\rho^2m_p^{-2}
\left(\Gamma(T,T_{wd})-\Lambda(T)\right).
\end{array}
\right. \label{HDC}
\end{equation}

Here, $\rho$ is the density, $\vec{v}=(u,v,w)$ the velocity vector, $P$ the
pressure, $\varepsilon$ the internal energy, $\Phi$ the Roche potential,
$m_p$ the proton mass, $\vec{\Omega}_{bin}$ the orbital angular velocity of
the system, and $\Gamma(T,T_{wd})$ and $\Lambda(T)$ the radiative heating and
cooling functions. The system of gas-dynamical equations was closed with the
ideal gas equation of state, $P=(\gamma-1)\rho\varepsilon$, where $\gamma$ is
the adiabatic index, which was specified to be~$\slantfrac53$.

This system of equations was solved numerically using the Roe-Osher 
method\cite{BinaryStars:2002,Roe:86,Chakravarthy-Osher:85}
adapted for a multiprocessor computer. A three-dimensional, noninertial,
Cartesian coordinate system corotating with the binary was used for the
modeling.

Only half of the space occupied by the disk was modeled, due to the symmetry
of the problem about the equatorial plane. The computational domain was
specified to contain both the disk and stream of matter from~$L_1$, and had a
size of~$1.12A \times 1.14A \times 0.17A$ (where $A$ is the binary separation).
The computational grid contained~$121\times 121\times 62$ cells and was
distributed among 81 processors in a twodimensional ($9\times 9$) matrix.
The grid became more dense closer to the equatorial plane, making it possible
to resolve the vertical structure of the disk.

The boundary conditions at the outer edges of the computational domain were
specified as follows. In all cells except those in the vicinity of~$L_1$, we
assumed a constant density, $\rho_{b}=10^{-8}\rho_{L_1}$, where~$\rho_{L_1}$
is the density at~$L_1$ , a temperature~$13600^\circ$~K, and zero velocity.
The accretor was specified as a hemisphere with a radius of~$10^{-2} A$.
Matter entering the cells forming the accretor was taken to fall onto the
star. The stream was defined as a boundary condition: matter with a
temperature of~$5800^\circ$~K and an $x$ velocity~$v_x=6.3$~km/s was injected
into a region with radius~$0.014 A$ centered on~$L_1$. The density of the
matter at~$L_1$ was specified so that the mass-transfer rate corresponded to
the observed rate in TV~Col. 

\begin{table}[b]
\begin{center}
\begin{tabular}{c|ccccc}
No. & $q$ & $M_1 (M_\odot)$ & $M_2 (M_\odot)$ & $A (R_\odot)$\\
\hline
\hline
1 & 0.3 & 1.4 & 0.42 & 1.923 \\
2 & 0.7 & 0.8 & 0.56  & 1.745 \\
3 & 0.93 & 0.6 & 0.56 & 1.655 \\
\end{tabular}
\end{center}
\caption{Model parameters $q$, $M_1$, $M_2$, $A$}\label{tbl1}
\end{table}

We made three computational runs with different parameters of the system.
Table~\ref{tbl1} gives the accretor~($M_1$) and donor~($M_2$) masses and the
binary separation~($A$) for the three models. The binary separation was
adjusted to be consistent with the observed orbital period of the TV~Col
system of 5.5~h. Since the component mass ratio for TV~Col is not known
precisely, we used the two extreme values $q=0.93$ and $q=0.3$ (which still
allow the disk to contain the 3:1 Lindblad resonance), as well as the
intermediate value $q=0.7$. For the given rate of mass inflow into the system,
the accretion rate in the model corresponded to~$\approx 10^{-8} M_{\odot}$.
The modeling covered large intervals of time (up to several dozen orbital
periods), making it possible to reach a quasistationary accretion regime with
a constant mass of the disk.

\section{Results of The Modeling}

\begin{figure}[t]
\begin{center}
\begin{tabular}{cc}
\epsfig{file=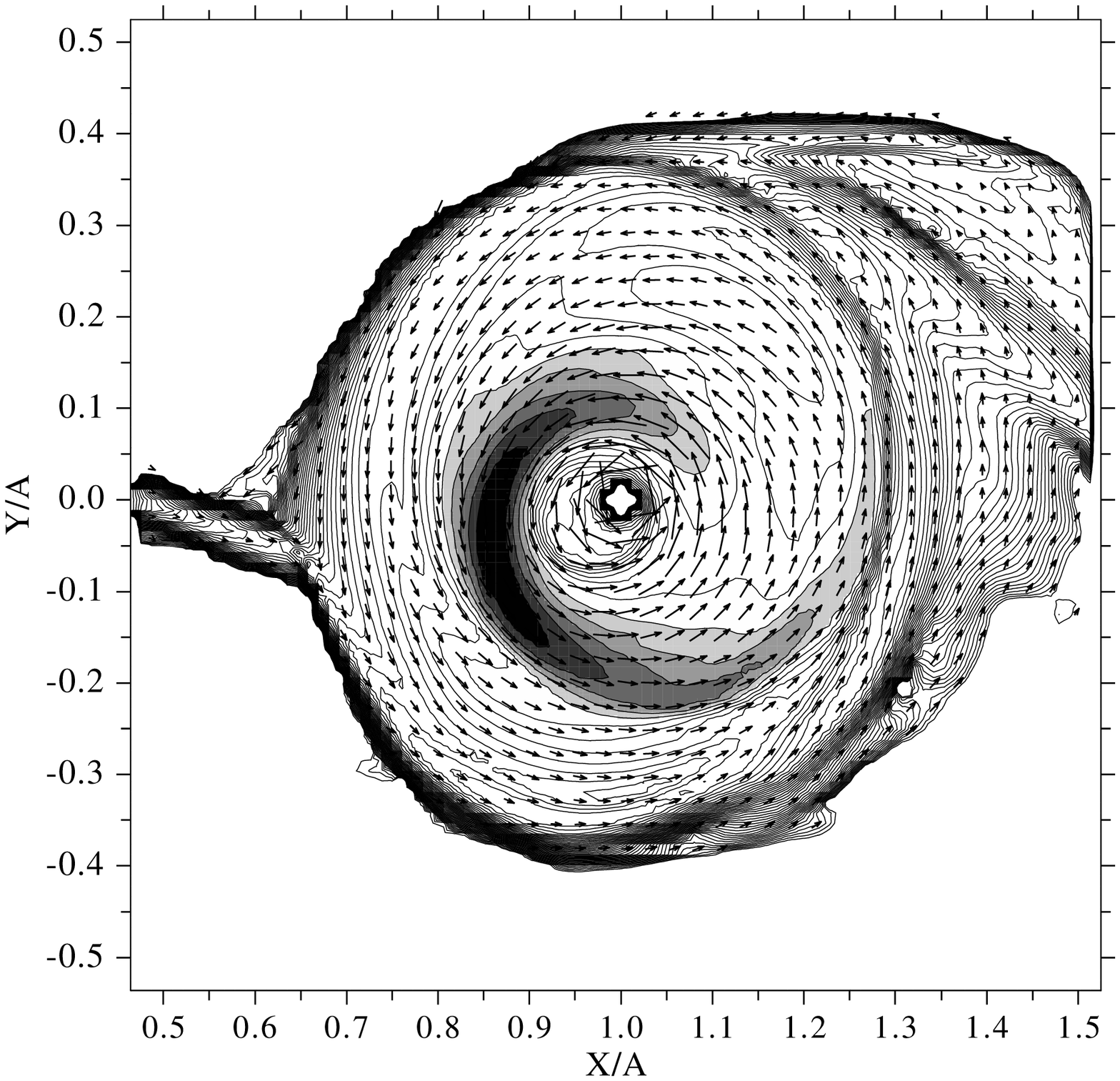, width=8cm}&
\epsfig{file=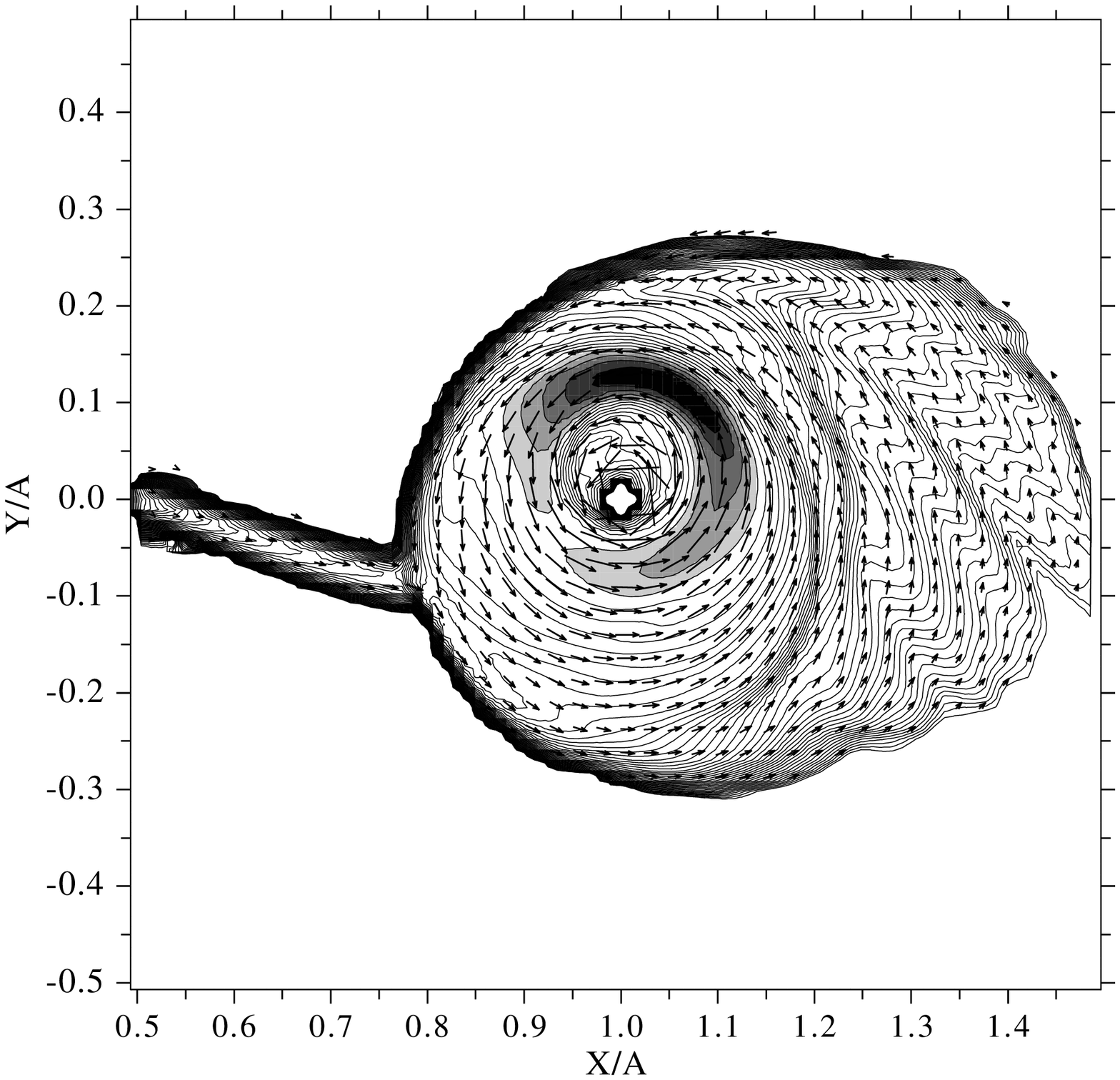,width=8cm}
\end{tabular}
\end{center}
\caption{
Density contours and velocity vectors in the equatorial plane of the system
in models with $q=0.7$ (left) and $q=0.93$ (right). The accretor is located
in the center of the region shown in the plot. In the vicinity of the
precessional wave, the density contours are shown using a gray scale. The
maximum density corresponds to black.
}\label{rhov}
\end{figure}

\begin{figure}[t]
\begin{center}
\begin{tabular}{cc}
\epsfig{file=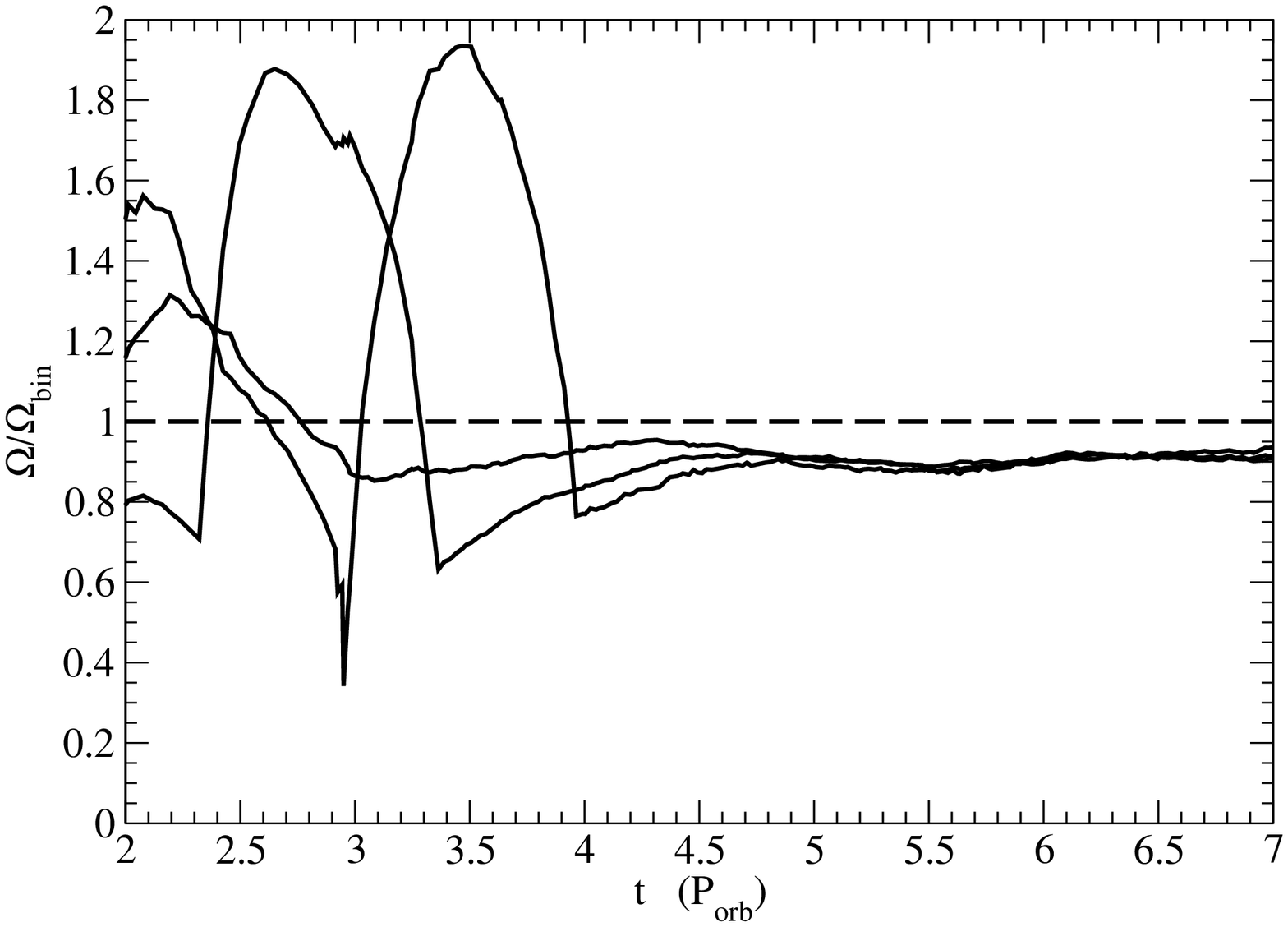, width=8cm}&
\epsfig{file=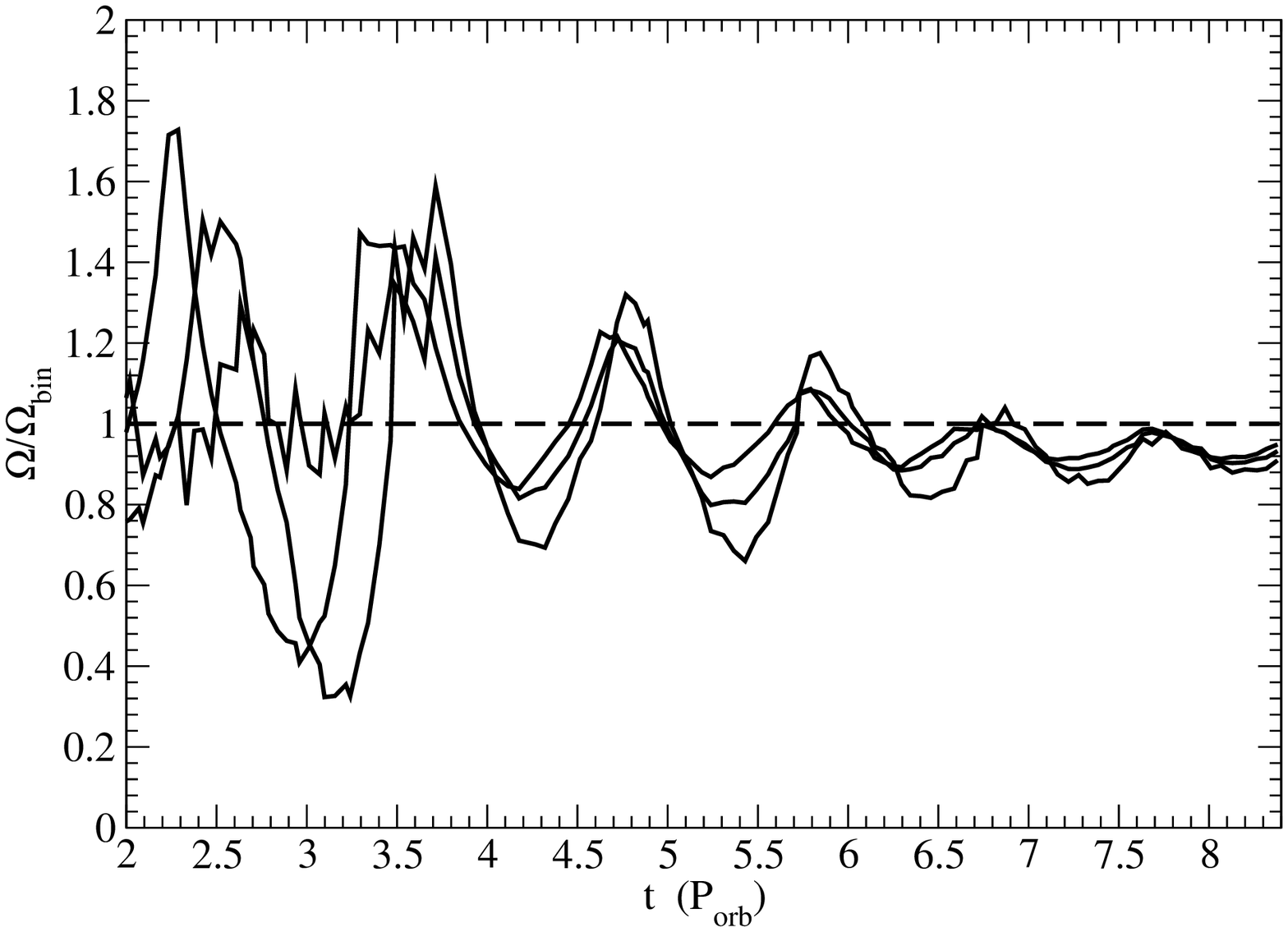,
width=8cm}
\end{tabular}
\end{center}
\caption{
Dependence of the angular velocity of the density peaks at three different
distances from the accretor in models with $q=0.7$ (left) and $q=0.93$
(right). Angular velocities are shown for a noninertial coordinate frame
rotating with the orbital angular velocity of the system. The horizontal
dashed line corresponds to the orbital velocity of the binary.
}\label{prec_spd}
\end{figure}

Figure~\ref{rhov} shows density contours and velocity vectors in the
equatorial plane for model nos. 2 and 3 from Table~\ref{tbl1}.

Places where the density contours become closer together correspond to
stationary shocks in the disk. A dense, round accretion disk and a compact
circumdisk halo form in the system. At the edge of the circumdisk halo, which
is formed by matter that collides with the stream when the former moves around
the accretor, the density rapidly falls to the background values. The
interaction of the circumdisk halo gas and the stream results in the formation
of a shock (hot line; see~\cite{AR:2000:Bisikalo-et-al,
BinaryStars:2002, AR:2003:CoolDiscsEng, AR:2004:PrecWave,
AR:2004:SUUMa, Bisikalo:2005, Bisikalo-et-al:ASP:2005,
Bisikalo-et-al:AIP:2005, AR:2005:Hump}) outside the disk. A twoarmed tidal
spiral shock forms in the disk, with both the arms located in the outer regions
of the disk. A precessional spiral density shock is clearly visible in the
inner regions of the disk. The two-armed spiral shock is at rest in the
noninertial coordinate frame fixed to the binary, while it moves with a slow
velocity in the inertial observer's frame. The period of the precessional wave
in the noninertial coordinate frame is slightly longer than the orbital period.

To find the rate of precession, we determined the phases at which the maximum
density is observed for various distances from the accretor. The precession
of the wave results in a shift of these phases with time. The typical time
dependences of the angular displacements of the density wave for three
distances from the accretor (the beginning, middle, and end of the wave) are
shown in Fig.~\ref{prec_spd}. These plots demonstrate that the velocities of
the streamlines in the disk level out, and the wave moves with some
equilibrium velocity $\Omega_{wav}$ in the coordinate frame fixed to the
binary.

\begin{table}[b]
\begin{center}
\begin{tabular}{c|ccc}
No. & $q$ & $\Omega_{wav}/\Omega_{bin}$ & $\epsilon^+$ \\
\hline
\hline
1 & 0.3 & -0.07 & 7 \% \\
2 & 0.7 & -0.1 & 11\% \\
3 & 0.93 & -0.05 & 5\% \\
\end{tabular}
\end{center}
\caption{
Rate of precession in the three models
}\label{tbl2}
\end{table}

Table~\ref{tbl2} lists the precession rates $\Omega_{wav}$ found via the
numerical modeling for systems with various component mass ratios~$q$.
Table~\ref{tbl2} also presents the
period excexxes~$\epsilon^+=\dfrac{P_{sh}^+-P_{orb}}{P_{orb}}$,
where~$P_{sh}^+$ is the superhump period.

\section{Connection Between Superhumps and Precessional Waves}

As a rule, five types of superhumps observed in cataclysmic variables can be
distinguished~\cite{ODonoghue:2000}.

(1) \underline{Positive} (or ordinary) superhumps have a period that is a few
percent
longer than the orbital period. Such superhumps were first observed in SU~UMa
during a superoutburst. 

(2) \underline{Orbital} superhumps represent a modulation of luminosity with
the orbital period.

(3) \underline{Late} superhumps are observed after the end of a superoutburst,
and have the same velocity as the positive superhump, but shifted by half a
period relative to the latter.

(4) \underline{Permanent} superhumps have the same features as positive
superhumps, but are observed in stars that lack superoutbursts.

(5) \underline{Negative} superhumps have a period that is several percent
shorter than the orbital period. Such superhumps were first discovered
during monitoring of systems with permanent superhumps.

The precessional spiral wave model is able to explain all types of observed
superhumps. The main observational features of positive and late superhumps
are a direct consequence of the formation of a precessional spiral wave in the
disk, as is explained in~\cite{AR:2004:SUUMa}. The orbital superhump can be
explained by the release of energy in stationary shocks in the disk and
circumdisk halo, such as the hot-line and tidal shocks; the existence of
tidal shocks is not precluded in the precessional wave model. Permanent
superhumps can also be explained, if the mass-transfer rate in the system is
high enough to sustain a high accretion rate due to prolonged existence of
the precessional spiral wave. The existence of luminosity modulations whose
period is shorter than the orbital period, i.e., of negative superhumps, can
also be plausibly explained in this model. 

\begin{figure}[t]
\begin{center}
\epsfig{file=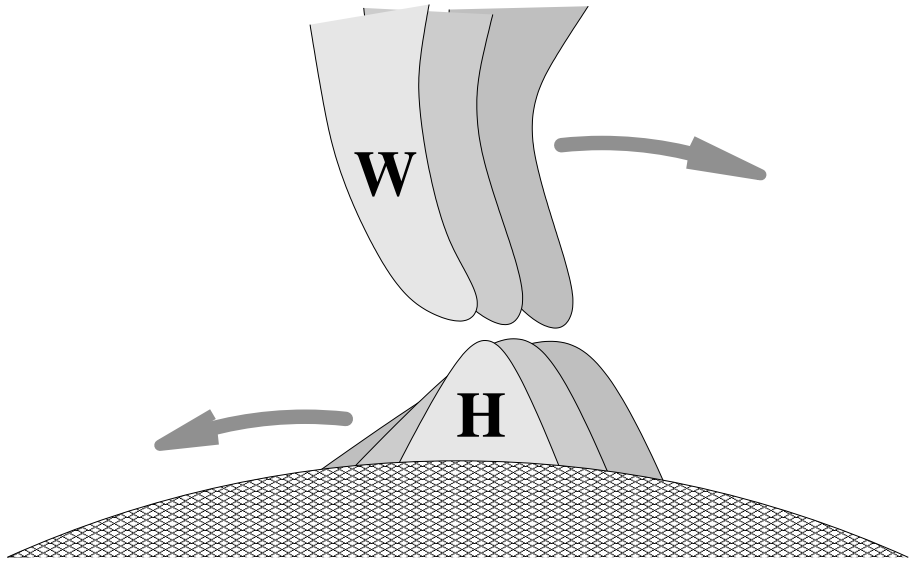, width=12cm}
\end{center}
\caption{
Formation of a negative superhump. The positions of the precessional wave (W)
and bright (hot) spot (H) are shown for three different times. The directions
in which the wave and leading edge of the superhump move are shown by arrows.
}\label{humps}
\end{figure}

In the model, the superhump radiation arises from a relatively compact region
near the surface of the accretor, into which matter flows along the
precessional wave. If the bright spot arising in this region is located above
the accretor surface (i.e., the rotation of accretor does not influence the
motion of the spot) and has a tendency to spread out due to diffusion, the
leading edge of the spot may be observed somewhat earlier after each rotation
of the system, creating a modulation with a period shorter than the orbital
period. The fact that the observed periods of negative superhumps do not
display significant scatter can be taken to justify the assumption that the
energy release occurs above the surface of the accretor. 

The necessary rate of diffusional spreading can be estimated using the
observed period. Let us denote the observed ``period excess'' for a negative
superhump~$\epsilon^-=\dfrac{P_{sh}^--P_{orb}}{P_{orb}}$, where~$P_{orb}$ is
the orbital period of the system and~$P^-_{sh}$ is the observed period of the
negative superhump. For TV~Col, $P^-_{sh}\sim
5.2$~h~\cite{Retter-et-al:2003}, so~$\epsilon^-\sim -0.055$. If the observed
period is the result that of beating between the orbital period and rotational
period of the leading edge of the spot~$P_{lead}$, this period can be found
using the formula 
$$
P_{lead}=\dfrac{P^-_{sh} P_{orb}}{P_{orb}-P^-_{sh}}=
-P_{orb}\dfrac{\epsilon^-+1}{\epsilon^-}\approx 95.3\,\,\text{h}.
$$
The linear velocity of the spot edge will be 
$$
v_{lead}=\dfrac{2\pi r_*}{P_{lead}}=
-\dfrac{2\pi r_*}{P_{orb}}\dfrac{\epsilon^-}{\epsilon^-+1}\approx
0.18\,\,\text{km/s},
$$
if the radius of the accretor is~$r_*\approx 10^9\,\,\text{cm}$. The velocity
of the leading edge has two components: the rate of diffusional spreading 
$v_{diff}$ and the (negative) velocity of the spot center, relative to the
retrograde precession of the wave, $v_{wav}$ . Knowing the period of the
precessional wave $P_{wav}$ (or the period excess of the positive superhump
$\epsilon^+$), we can find the linear velocity of the wave, $v_{wav}$:
$$
v_{wav}=-\dfrac{2\pi r_*}{P_{wav}}=
-2\pi r_*\dfrac{P^+_{sh}-P_{orb}}{P^+_{sh}P_{orb}}=
-\dfrac{2\pi r_*}{P_{orb}}\cdot\dfrac{\epsilon^+}{\epsilon^++1}\approx
-0.4\,\,\text{km/s}.
$$
This velocity is negative, since the direction of the wave precession is
retrograde. Accordingly, $v_{diff}$ will be given by
$$
v_{diff}=v_{lead}-v_{wav}\approx 0.58\,\,\text{km/s}.
$$

Let find a relation between the diffusion rate derived from observations and
the parameters of the disk gas. The time scale for diffusion over a distance
$L$ is~$\tau=L^2/D$, where $D$ is the diffusion coefficient~\cite{Lang:74}.
If~$\tau=P_{orb}$, the linear diffusion rate is 
$$
v_{diff}=\dfrac{L}{\tau}=\sqrt{\dfrac{D}{P_{orb}}}.
$$
In this case, the diffusion coefficient for the TV~Col system will be
$$
D=P_{orb}\cdot v^2_{diff}\sim\dfrac{1.1\cdot10^{16}}{P_{orb}}\left(
-\dfrac{\epsilon^-}{1+\epsilon^-}+\dfrac{\epsilon^+}{1+\epsilon^+}
\right)^2\approx 6.8\cdot10^{13}\,\,\text{cm}^2/\text{s}
$$

The diffusion coefficient for an ionized gas is defined as $D=l\cdot
V_{rms}$, and depends on the mean free path $l$ and rms velocity $V_{rms}$.
For ionized gas, $l=\dfrac{3.2\cdot 10^6\cdot T^2}{n\,\mathrm{ln}\Lambda}$,
where $T$ is the temperature of the gas, $n$ the number density of particles,
and $\Lambda=\dfrac{1.3\cdot 10^4 T^{3/2}}{\sqrt{n}}$. The rms velocity is
$\left(\dfrac{3kT}{m}\right)^{1/2}$, where $k$ is Boltzmann's constant and~$m$
is the mass of the gas particles. Accordingly, the diffusion coefficient is a
function of the temperature and number density of the matter: $D=f(n,T)$.
Thus, if we know the value of $D$ from observations, it is possible to find
$n$ and $T$.

\begin{figure}[t]
\begin{center}
\epsfig{file=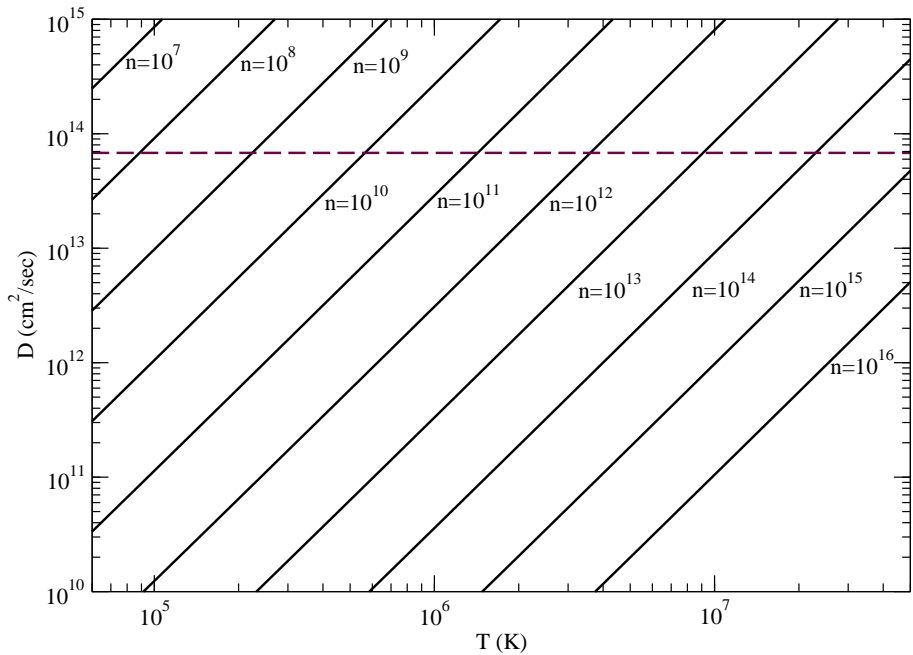, width=12cm}
\end{center}
\caption{
Dependence $D(T)$ for various $n$. The dashed line corresponds to the $D$
observed in TV~Col.
}\label{D_nt}
\end{figure}

\begin{figure}[t]
\begin{center}
\epsfig{file=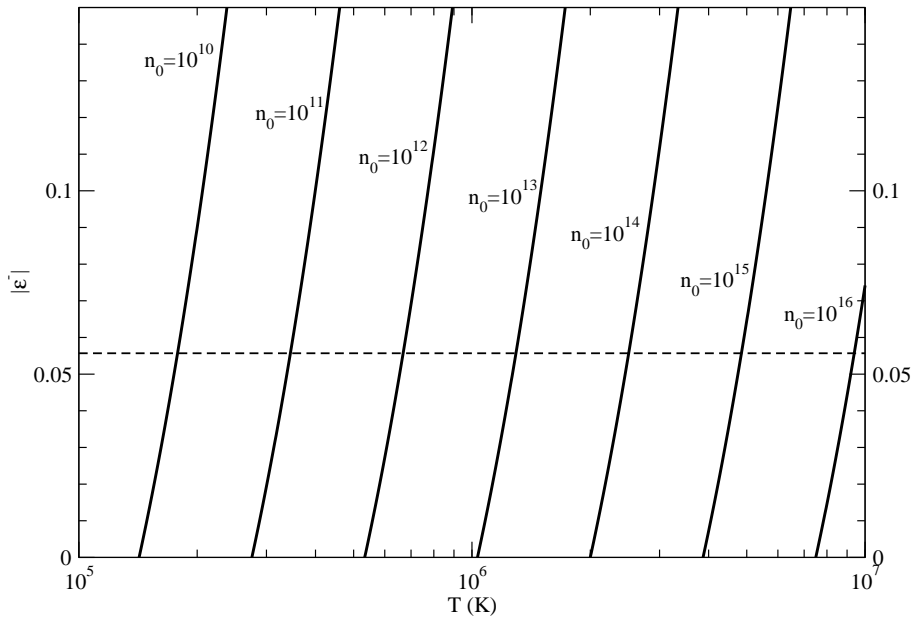, width=12cm}
\end{center}
\caption{
Theoretical dependence $|\epsilon^-(T)|$ in TV~Col for various values of
$n_0$. The dotted line corresponds to the $|\epsilon^-|$ observed in TV~Col.
}\label{eps_t}
\end{figure}

Figure~\ref{D_nt} shows the dependence~$D(T)$ for various $n$.
The dashed line shows the value of~$D$ that corresponds to the observed
diffusion for TV~Col. If we can estimate the typical temperature from
observations of the superhump, Fig.~\ref{D_nt} can be used to find
the density of matter in the spot. Let us now assume that there is pressure
equilibrium in the accretion disk, so that $nT=n_{0}T_{0}$, where $n_0$ and
$T_0$ are the number density and temperature of the outer parts of the disk.
Assuming that~$T_0\sim 10^4 K$, we can find the typical number density in the
disk, $n_0$. Figure~\ref{eps_t} shows a set of solutions for various $n_0$ 
(for $T_0=10^4 K$). For convenience, we also show~$|\epsilon^-(T)|$ for TV~Col
in this figure. If we suppose the temperature of the spot to be $T\sim 10^6
K$ and use the observed value of~$|\epsilon^-|$ (dashed line), then
$n_0\approx 10^{13} cm^{-3}$. The estimated values of $T$ and $n_0$ are within
the range typical for accretion disks in close binary
systems~\cite{AR:2003:CoolDiscsEng}. Thus, this model with the diffusional
spreading of the spot can adequately explain the observed negative superhumps.

\section{Conclusions}

Our numerical modeling has shown that precessional density waves can form in
the accretion disks of systems with large component mass ratios, up to 
$q=0.93$.

\begin{figure}[t]
\begin{center}
\epsfig{file=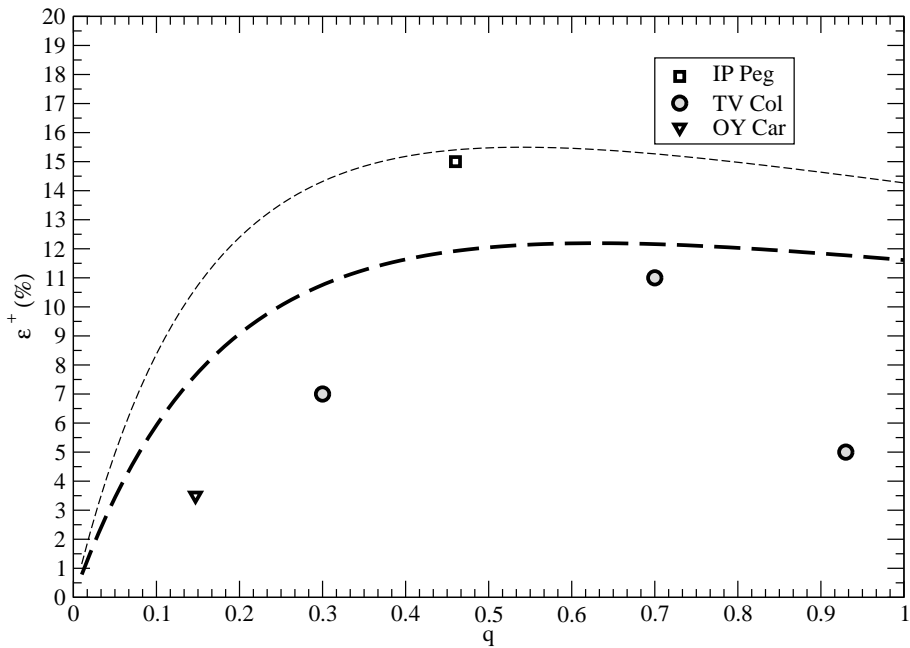, width=12cm}
\end{center}
\caption{
Dependence of the superhump period excess on the component mass ratio. The
bold dashed line corresponds to the theoretical dependence given
in~\cite{Hirose-Osaki:90}. The thin dashed line corresponds to the same
dependence for a disk whose radius is 10\% larger. The open circles show the
period excess of the precessional density wave we have found for various
possible values of $q$ for TV~Col. The triangle corresponds to the period
excess found for OY~Car in~\cite{AR:2004:SUUMa}, and the square to the period
excess found for IP~Peg in~\cite{AR:2004:PrecWave}.
}\label{prec}
\end{figure}

Our derived dependence of the precession period on~$q$ is complex.
Figure~\ref{prec} shows the theoretical dependence~$\epsilon^+(q)$ obtained
in~\cite{Hirose-Osaki:90} (bold dashed line). The open circles in this figure
show the period excesses found for three different sets of parameters of 
TV~Col. We also show the model~$\epsilon^+(q)$ values for OY~Car and IP~Peg
found in~\cite{AR:2004:SUUMa,AR:2004:PrecWave}. Despite the significant
scatter, all the~$\epsilon^+(q)$ values are fairly close to the theoretical
line, although the values for $q=0.49$~(IP~Peg) and $q=0.93$~(TV~Col) are
outliers. Note that the rotational rate of the precessional density wave
depends not only on the component mass ratio, but also on the size of the
region occupied by the wave. The wave can exist only in a gas-dynamically
unperturbed region of the disk, whose size is limited by the depth of
penetration of stationary shocks -- the two arms of the tidal shocks and the
hot line (extended region of interaction of the stream from $L_1$ and the
circumdisk halo) -- into the disk. The high masstransfer rate in TV~Col may
result in an increase in the extent of the hot line and, consequently, to
precession rates that are lower than expected theoretically. It is interesting
that, in the case of high component mass ratios, the size of the unperturbed
region may be significantly limited by tidal waves, which would result in an
even larger decrease of the precession rate with increasing $q$, compared to
the theoretical value. Nevertheless, the dependence is close to the
theoretical dependence, and has a peak in the range $q\sim 0.4\div 0.6$.

In TV~Col, superhumps with the period 
$P_{sh}\sim 6.3$~h~\cite{Retter-et-al:2003} are observed, so that the
superhump period excess is $\sim 15\%$. This value is out of the range of the
theoretical estimates of the precession rate (but does not exceed the value
obtained in~\cite{AR:2004:PrecWave} for IP~Peg). The theoretical dependence
for~$\epsilon^+(q)$ was obtained in the model that did not take into account
the gas pressure; i.e., it effectively estimates the precession rate of the
semimajor axis of the orbit of a free particle. The excesses of the observed
and simulated precession rates over the theoretical values may indicate that
the estimated disk radius used in the simulations,
$r_{disc}=\dfrac{0.6}{1+q}$~\cite{Paczynski:77}, is too low (in fact, the
derivations in~\cite{Paczynski:77} did not take into account the gas pressure).
If the disk radius were 10\% larger ($r_{disc}=\dfrac{0.66}{1+q}$), all the
derived precession rates would be below the theoretical line (shown by the
thin dashed line in Fig.~\ref{prec}).

In summary, we conclude that the presence of a precessional density wave in
the disk can lead to superhumps both in SU~UMa stars and in binaries whose
component mass ratios prohibit the location of the 3:1 Lindblad resonance
inside the disk. The precessional wave model is able to explain all types of
observed superhumps.

\section*{Acknowledgments}

This work was supported by the Russian Foundation for Basic Research (project
nos. 05-02-16123, 05-02-17070, 05-02-17874, 06-02-16097, and 06-02-16234),
the Program of Support for Leading Scientific Schools of Russia, and the
Basic Research Programs of the Presidium of the Russian Academy of Sciences 
``Origin and Evolution of Stars and Galaxies'', ``Basic Problems in
Informatics and Informational Technologies", the Programm of Support for 
Young Scientists of Russia and the Russian Science Support Foundation.


\end{document}